\documentclass[dvips,12pt,a4paper]{article}
\usepackage{a4wide}
\usepackage{epsfig}
\usepackage{amsmath}
\usepackage{latexsym}

  \newlength{\absize}
\catcode`@=11
\def\citer{\@ifnextchar [{\@tempswatrue\@citexr}{\@tempswafalse\@citexr[]}}
 
\def\@citexr[#1]#2{\if@filesw\immediate\write\@auxout{\string\citation{#2}}\fi
  \def\@citea{}\@cite{\@for\@citeb:=#2\do
    {\@citea\def\@citea{--\penalty\@m}\@ifundefined
       {b@\@citeb}{{\bf ?}\@warning
       {Citation `\@citeb' on page \thepage \space undefined}}%
\hbox{\csname b@\@citeb\endcsname}}}{#1}}
\catcode`@=12

\begin{document}
  \thispagestyle{empty}
  \pagestyle{empty}
  \renewcommand{\thefootnote}{\fnsymbol{footnote}}
\newpage\normalsize
    \pagestyle{plain}
    \setlength{\baselineskip}{4ex}\par
    \setcounter{footnote}{0}
    \renewcommand{\thefootnote}{\arabic{footnote}}
\newcommand{\preprint}[1]{%
  \begin{flushright}
    \setlength{\baselineskip}{3ex} #1
  \end{flushright}}
\renewcommand{\title}[1]{%
  \begin{center}
    \LARGE #1
  \end{center}\par}
\renewcommand{\author}[1]{%
  \vspace{2ex}
  {\Large
   \begin{center}
     \setlength{\baselineskip}{3ex} #1 \par
   \end{center}}}
\renewcommand{\thanks}[1]{\footnote{#1}}
\renewcommand{\abstract}[1]{%
  \vspace{2ex}
  \normalsize
  \begin{center}
    \centerline{\bf Abstract}\par
    \vspace{2ex}
    \parbox{\absize}{#1\setlength{\baselineskip}{2.5ex}\par}
  \end{center}}

\begin{flushright}
{UTS-DFT-99-01}           \\
\end{flushright}
\vspace*{4mm}
\vfill
\title{Optimal polarized observables for model-independent new physics searches 
at the linear collider\footnote{Contribution to the International Workshop on
Linear Colliders LCWS99, Sitges (Barcelona), Spain, 28 Apr-5 May 1999} }
\vfill
\author{N. Paver \footnote{Partially supported by MURST (Italian Ministry of 
University, Scientific Research and Technology)}}
\begin{center}
Dipartimento di Fisica Teorica, Universit\`a di Trieste and \\
Istituto Nazionale di Fisica Nucleare, Sezione di Trieste,
Trieste, Italy
\end{center}
\vfill
\abstract
{We discuss the sensitivity to four-fermion 
contact interactions of the process of fermion-pair production at the $e^+-e^-$ 
Linear Collider (LC), with $E_{CM}=0.5\ {\rm TeV}$ and 
longitudinally polarized electron beam. The analysis is based on polarized  
integrated cross sections with optimal cuts. Beam polarization and optimization 
are shown to have a crucial role in the derivation of model-independent 
constraints on the new interactions from the data.} 
\vspace*{20mm}
\setcounter{footnote}{0}

\newpage
    \setcounter{footnote}{0}
    \renewcommand{\thefootnote}{\arabic{footnote}}
    \setcounter{page}{1}

\section{Introduction and polarized cross sections}

The concept of effective contact interaction represents a convenient framework
to parameterize physical effects of some new dynamics active at a very high  
scale $\Lambda$, in reactions among the `light', Standard Model, degrees of 
freedom such as quarks and leptons, $W$, $Z$, etc., at `low energy' 
$E\ll\Lambda$. These effects are suppressed by an inverse power of the large 
scale $\Lambda$, and should manifest by deviations of experimentally measured  
observables from the SM predictions. We consider here the 
process, at LC energy: 
\begin{equation}
e^++e^-\to\bar{f}+f, \label{eq:proc}
\end{equation}
and the relevant $SU(3)\times SU(2)\times U(1)$ symmetric, lowest-dimensional 
$eeff$ contact-interaction Lagrangian with helicity conserving, flavor-diagonal 
fermion currents \cite{eichten}:
\begin{equation}
{{\cal L}}=\sum_{\alpha,\beta}\frac{g^2_{\rm eff}}{\Lambda^2_{\alpha\beta}}
\eta_{\alpha\beta}\left(\bar e_{\alpha}\gamma_\mu e_{\alpha}\right)
\left(\bar f_{\beta}\gamma^\mu f_{\beta}\right).
\label{eq:lagra}
\end{equation}
In Eq.~(\ref{eq:lagra}), generation and color indices have been suppressed, 
$\alpha,\beta=L,R$ indicate left- or right-handed helicities, and the 
parameters $\eta_{\alpha\beta}=\pm 1,0$ specify the independent, 
individual, interaction models. Conventionally, $g^2_{\rm eff}=4\pi$ as a
reminder that the new interaction, originally proposed for compositeness, 
would become strong at $E\sim\Lambda$. Thus, in practice, the scales 
$\Lambda_{\alpha\beta}$ define a standard to compare the reach of different 
new-physics searches. For example, a bound on $\Lambda$ in the case of a very 
heavy $Z^\prime$ exchange with couplings of the order of the electron charge 
would translate into a constraint on the mass 
$M_{Z^\prime}\sim{\sqrt\alpha}\Lambda$, and the same is true for leptoquarks or 
for any other heavy object exchanged in 
process (\ref{eq:proc}) \cite{sriemann}.     

Constraints on $\cal L$ can be obtained by looking at deviations of observables
from the SM predictions in the experimental data. In general, such 
sought-for deviations can simultaneously depend on all four-fermion effective 
coupling constants, that cannot be easily disentangled. A commonly adopted
possibility is to assume non-zero value for only one parameter at a time, 
with the remaining ones set equal to zero. In this way, one would test the
individual models mentioned above, and current bounds from a global analysis of
the relevant data are of the order of 
$\Lambda_{\alpha\beta}\sim{\cal O}(10)\ {\rm TeV}$ \cite{zeppenfeld}. However, 
for the derivation 
of model-independent constraints, a procedure that allows to simultaneously 
take into account the terms of different chiralities, and at the same time to 
disentangle the contributions of the different individual couplings to avoid 
potential cancellations that can weaken the bounds, is highly desirable. To
this purpose, initial beam longitudinal polarization would offer the 
possibility of defining polarized cross sections, that allow to 
reconstruct from the data the helicity cross sections depending on the 
individual $eeff$ chiral couplings of Eq.~(\ref{eq:lagra}), and consequently to 
make an analysis in terms of a minimal set of free independent 
parameters \cite{pankov,cheung}. Also, integrated cross sections 
would be of advantage in the case of limited statistics, and some optimal
choice of the kinematical region may further improve the sensitivity to the new
interaction. 

For $f\neq e,t$ and $m_f\ll\sqrt s\equiv E_{CM}$, the differential cross 
section for process (\ref{eq:proc}) is determined in Born approximation by the 
s-channel $\gamma, Z$ exchanges plus ${\cal L}$ of Eq.~(\ref{eq:lagra}). With 
$P_e, P_{\bar e}$ the initial beams longitudinal 
polarization $\hskip 2pt$\cite{schrempp}:  
\begin{equation}
\frac{d\sigma}{d\cos\theta}
=\frac{3}{8}
\left[(1+\cos\theta)^2\hskip 2pt {\tilde\sigma}_+ 
+(1-\cos\theta)^2\hskip 2pt {\tilde\sigma}_-\right], 
\label{eq:cross}
\end{equation}
where, in terms of helicity cross sections 
\begin{eqnarray}
{\tilde\sigma}_{+}&=&\frac{1}{4}\,
\left[(1-P_e)(1+P_{\bar{e}})\,\sigma_{\rm LL}
+(1+P_e)(1- P_{\bar{e}})\,\sigma_{\rm RR}\right], \label{eq:s+} \\
{\tilde\sigma}_{-}&=&\frac{1}{4}\,
\left[(1-P_e)(1+ P_{\bar{e}})\,\sigma_{\rm LR}
+(1+P_e)(1-P_{\bar{e}})\,\sigma_{\rm RL}\right], \label{eq:s-} 
\end{eqnarray}
and ($\alpha,\beta=L,R$; $N_C\approx 3(1+\alpha_s/\pi)$ for quarks and $N_C=1$ 
for leptons):
\begin{equation} 
\sigma_{\alpha\beta}=N_C\frac{4\pi\alpha^2_{em}}{3s}
\vert A_{\alpha\beta}\vert^2.
\label{eq:helcross}
\end{equation}
The helicity amplitudes are 
\begin{equation}
A_{\alpha\beta}=Q_e\hskip 1pt Q_f+g_\alpha^e\,g_\beta^f\,\chi_Z+
\frac{s\eta_{\alpha\beta}}{\alpha\Lambda_{\alpha\beta}^2}, 
\label{eq:amplit} 
\end{equation}
where $Q$'s and $g$'s are are the fermion electric charges and SM chiral 
couplings, respectively, and $\chi_Z=s/(s-M_Z^2+is\Gamma_Z/M_Z)$.

The above relations clearly show that the helicity cross sections, that 
directly relate to the individual four-fermion contact interaction
couplings and therefore allow a model-independent analysis, can be 
disentangled by the measurement of ${\tilde\sigma}_{+}$ and 
${\tilde\sigma}_{-}$ with different choices of the initial beam polarizations, 
and making linear combinations. 
In particular, one can easily see by integration of Eq.~(\ref{eq:cross}) in 
$\cos\theta$ that the `conventional' observables, the total cross section 
$\sigma\equiv\sigma_F+\sigma_B={\tilde\sigma}_{+}+{\tilde\sigma}_{-}$ and the 
forward-backward difference $\sigma_{FB}\equiv\sigma_F-\sigma_B
=\frac{3}{4}\left({\tilde\sigma}_{+}-{\tilde\sigma}_{-}\right)$, depend an all 
helicity cross sections and therefore do not allow the separation by
themselves, unless their measurements at different initial polarizations are 
combined (a minimum of four measurements is needed).

For the discussion of the expected uncertainties and the corresponding
sensitivities to the parameters of $\cal L$, as well as for improving the
significance of the resulting bounds on $\Lambda_{\alpha\beta}$, one can more 
generally define polarized cross sections integrated over the {\it a priori} 
arbitrary kinematical ranges $(-1,z^*)$ and $(z^*,1)$: 
\begin{eqnarray} 
\label{eq:sigma1}
\sigma_1(z^*)
&\equiv&\int_{z^*}^1\frac{d\sigma}{d\cos\theta}d\cos\theta
=\frac{1}{8}\left\{\left[8-(1+z^*)^3\right]\tilde\sigma_++(1-z^*)^3
\hskip 2pt\tilde\sigma_-\right\}, \\
\label{eq:sigma2}
\sigma_2(z^*)
&\equiv&\int^{z^*}_{-1}\frac{d\sigma}{d\cos\theta}d\cos\theta
=\frac{1}{8}\left\{(1+z^*)^3\hskip 2pt\tilde\sigma_++
\left[8-(1-z^*)^3\right]\tilde\sigma_- 
\right\},
\end{eqnarray}
and try to disentangle the helicity cross sections from the general relations,
at different values of the polarizations $P_e$ and $P_{\bar e}$: 
\begin{eqnarray}
\label{eq:sigmap}
\tilde\sigma_+
&=&\frac{1}{6(1-{z^*}^2)}\left[\left(8-(1-z^*)^3\right)
\hskip 2pt\sigma_1(z^*)-(1-z^*)^3\hskip 2pt\sigma_2(z^*)\right], \\
\label{eq:sigmam}
\tilde\sigma_-
&=&\frac{1}{6(1-{z^*}^2)}\left[-(1+z^*)^3\hskip 2pt\sigma_1(z^*)
+\left(8-(1+z^*)^3\right)\hskip 2pt\sigma_2(z^*)\right].
\end{eqnarray}                                      
In practice, we adopt $P_e=\pm P$ with $P < 1$ and $P_{\bar e}=0$. Then, the
basic set of integrated observables are $\sigma_{1,2}(z^*,P_e)$ and, as a
second step, we construct the cross sections ${\tilde\sigma}_\pm(P_e)$ which 
finally yield the helicity cross sections $\sigma_{\alpha\beta}$  by solving 
the linear system of equations corresponding to the two signs of $P_e$. 

One can easily see that the specific choice $z^*=0$ in Eqs.~(\ref{eq:sigma1}) 
and (\ref{eq:sigma2}) leads back to the forward and backward cross sections 
$\sigma_F$ and $\sigma_B$.
Instead, the values $z^*=z^*_\pm=\mp(2^{2/3}-1)=\mp 0.587$ 
($\theta^*_+=126^\circ$ and $\theta^*_-=54^\circ$) to a very good approximation 
allow to directly `project' out 
${\tilde\sigma}_\pm$ \cite{pankov,osland}: 
\begin{equation}
\label{eq:tildesigmas}
{\tilde\sigma}_{+}=\gamma\left(\sigma_{1}(z^*_+)-\sigma_{2}(z^*_+)\right), 
\qquad
{\tilde\sigma}_{-}=\gamma\left(\sigma_{2}(z^*_-)-\sigma_{1}(z^*_-)\right),
\end{equation}
where $\gamma=[3\hskip1pt\left(2^{2/3}-2^{1/3}\right)]^{-1}=1.018$.
Finally, $z^*$ could be taken as an input parameter related to given
experimental conditions, that can be tuned to get maximal sensitivity of
$\sigma_{\alpha\beta}$ on $\Lambda$'s \cite{pankov1}.  
 
\section{Numerical analysis, optimization and bounds on $\Lambda$}

We adopt a $\chi^2$ procedure, defined as follows:
\begin{equation}
\chi^2=
\left(\frac{\Delta\sigma_{\alpha\beta}}{\delta\sigma_{\alpha\beta}}\right)^2,
\label{eq:chi2}
\end{equation}
where
$\Delta\sigma_{\alpha\beta}=\sigma_{\alpha\beta}-\sigma_{\alpha\beta}^{SM}$ 
are the deviations of helicity cross 
sections due to the contact four-fermion interaction (\ref{eq:lagra}), and 
$\delta\sigma_{\alpha\beta}$ are the 
corresponding expected experimental uncertainties on $\sigma_{\alpha\beta}$, 
combining both statistical and systematic uncertainties. Assuming that no 
deviation from the SM is observed within the experimental accuracy, 
constraints on the allowed values of $\Lambda$'s are obtained by imposing 
$\chi^2 < \chi^2_{\rm crit}$, where the actual value of $\chi^2_{\rm crit}$ 
specifies the desired `confidence level' ($\chi^2_{\rm crit}= 3.84$ as typical 
for 95\%  C.L. with a one-parameter fit). For the expected uncertainties on 
$\sigma_{1,2}$, we assume the following identification efficiencies 
($\epsilon$) and systematic uncertainties ($\delta^{\rm sys}$) for the 
different final states \cite{damerell}: 
$\epsilon=100\%$ and $\delta^{\rm sys}=0.5\%$ for leptons; 
$\epsilon=60\%$ and $\delta^{\rm sys}=1\%$ for $b$ quarks; $\epsilon=35\%$ and 
$\delta^{\rm sys}=1.5\%$ for $c$ quarks. To have an indication on the role of
statistics, for the LC with $\sqrt{s}=0.5\ {\rm TeV}$ we consider  
time-integrated total luminosities $L_{\rm int}=50$ and 500 $\mbox{fb}^{-1}$, 
and assume $1/2\hskip 2pt L_{\rm int}$ for each values $P_e=\pm P$. We take the 
values $P=$ 1, 0.8, 0.5 as a reasonable variation around $P=0.8$, expected at 
the LC \cite{accomando}, in order to study the dependence of the results on the 
initial beam longitudinal polarization. The numerical analysis uses the program 
ZFITTER \cite{bardin} along with ZEFIT, with input values 
$m_{\rm top}=175\ {\rm GeV}$ and $m_H=100\ {\rm GeV}$. It takes into account 
one-loop SM electroweak corrections in the form of improved Born 
amplitudes \cite{consoli}, as well as initial- and final-state radiation with a 
cut on the photon energy emitted in the initial state 
$\Delta=E_\gamma/E_{\rm beam}=0.9$ to avoid radiative return to the $Z$ peak. 

The results for the reachable mass scales $\Lambda$ are shown in Tables~1
and 2 \cite{pankov1}. The left entries in each box represent the values 
obtained by the 
polarized integrated cross sections defined by $z^*_\pm$, see 
Eq.~(\ref{eq:tildesigmas}).  
\begin{table}[ht]
\centering
\caption{Contact-interaction reach (in TeV) for luminosity 50 ${\rm fb}^{-1}$, 
at 95\% C.L. The arrows indicate the increase of sensitivity of the 
observables caused by the optimization.}
\medskip
\begin{tabular}{|c|c|c|c|c|c|} 
\hline
&&&&& \\
process & $P$ & $\Lambda_{LL}$ & $\Lambda_{RR}$ & 
$\Lambda_{LR}$ & $\Lambda_{RL}$ \\
&&&&& \\
\hline
\cline{2-6}
 & 1.0 &$40\rightarrow 41$&$39\rightarrow 40$&
$26\rightarrow 40$&$28\rightarrow 41$ \\ 
\cline{2-6}
$\mu^+\mu^-$ & 0.8 &$37\rightarrow 38$&$37\rightarrow 38$&
$25\rightarrow 37$&$26\rightarrow  37$
\\ \cline{2-6}
 & 0.5 &$32\rightarrow 32$&$31\rightarrow 32$&
$21\rightarrow 30$&$21\rightarrow 30$
\\ \hline
\hline
 &1.0&$41\rightarrow 42$&$45\rightarrow 47$&
$17\rightarrow 31$&$34\rightarrow 42$
\\ \cline{2-6}
${\overline{b}}b$ &0.8&$40\rightarrow 41$&$38\rightarrow 39$&
$17\rightarrow 29$&$29\rightarrow 38$
\\ \cline{2-6}
&0.5&$36\rightarrow 37$&$29\rightarrow 29$&
$13\rightarrow 25$&$22\rightarrow 31$
\\ \hline
\hline
 &1.0&$32\rightarrow 33$&$36\rightarrow 37$&
$21\rightarrow 32$&$20\rightarrow 30$
\\ \cline{2-6}
${\overline{c}}c$ &0.8&$31\rightarrow 32$&$32\rightarrow 33 $&
$20\rightarrow 31$&$18\rightarrow 27$
\\ \cline{2-6}
 &0.5&$27\rightarrow 28$&$26\rightarrow 27$&
$18\rightarrow 27$&$15\rightarrow 22$
\\ \hline
\end{tabular}
\label{tab:tab1}
\end{table}
As one can see, the best sensitivity occurs for
$b\bar b$ production while the worst one is for $c\bar c$, and the decrease of
electron polarization $P$ from 1 to 0.5 worsens the sensitivity by 20-40\%, 
depending on the final state. As regards the role of the luminosity, the bounds
on $\Lambda$ would scale like $(L_{\rm int})^{1/4}$ if no systematic
uncertainty were assumed\cite{schrempp}, giving a factor 1.8 of improvement 
from 50 to 500 ${\rm fb}^{-1}$. This is the case of $\Lambda_{RL}$ and 
$\Lambda_{LR}$, where the dominant uncertainty is the statistical one, whereas 
the bounds for $\Lambda_{LL}$ and $\Lambda_{RR}$ depend much more sensitively 
on $\delta^{\rm sys}$. Moreover, it should be noticed that the sensitivity of
$\sigma_{RL}$ and $\sigma_{LR}$ is considerably smaller than that of 
$\sigma_{LL}$ and $\sigma_{RR}$. Thus, it is important to construct optimal 
obervables to get the maximum sensitivity.    
\begin{table}[t]
\centering
\caption{Same as Table~1, but at luminosity 500 ${\rm fb}^{-1}$.}
\medskip
\begin{tabular}{|c|c|c|c|c|c|} 
\hline 
&&&&& \\
process & $P$ & 
$\Lambda_{LL}$ & $\Lambda_{ RR}$ & 
$\Lambda_{LR}$ & $\Lambda_{ RL}$ \\
&&&&&\\ 
\hline
\cline{2-6}
 & 1.0 &$54\rightarrow 55$&$55\rightarrow 56$&
$37\rightarrow 57$&$40\rightarrow 59$ \\ 
\cline{2-6}
$\mu^+\mu^-$ & 0.8 &$51\rightarrow 52$&$51\rightarrow 53$&
$35\rightarrow  54$&$38\rightarrow 55$
\\ \cline{2-6}
& 0.5 &$44 \rightarrow 45$&$43 \rightarrow 44$&
$31\rightarrow 46$&$31\rightarrow 47$ \\ \hline
\hline
 & 1.0 &$48\rightarrow 49$&$64\rightarrow 67$&
$26\rightarrow 50$&$50\rightarrow 66$\\ 
\cline{2-6}
$\bar{b}b$ & 0.8 &$47\rightarrow 48$&$51\rightarrow 53$&
$25\rightarrow  48$&$41\rightarrow 61$
\\ \cline{2-6}
& 0.5 &$43\rightarrow 44$&$36 \rightarrow 37$&
$23\rightarrow 43$&$30\rightarrow 47$ \\ \hline
\hline
 & 1.0 &$35\rightarrow 37$&$42\rightarrow 43$&
$24\rightarrow 44$&$24\rightarrow 47$\\ 
\cline{2-6}
$\bar{c}c$ & 0.8 &$34\rightarrow 35$&$38\rightarrow 39$&
$23\rightarrow  43$&$22\rightarrow 41$
\\ \cline{2-6}
& 0.5 &$30\rightarrow 32$&$29 \rightarrow 30$&
$21\rightarrow 39$&$18\rightarrow 32$ \\ \hline
\end{tabular}
\label{tab:tab2}
\end{table}
Referring to Eq.~(\ref{eq:chi2}), the uncertainties 
$\delta\sigma_{\alpha\beta}$ depend on $z^*$ trough 
Eqs.~(\ref{eq:sigma1})-(\ref{eq:sigmam}), while $\Delta\sigma_{\alpha\beta}$ 
are $z^*$-independent. This $z^*$ dependence determines the sensitivity of 
each helicity amplitude to the corresponding $\Lambda$. It can be explicitly 
evaluated, for the statistical uncertainty, from the known SM cross sections 
and $L_{\rm int}$. Then, 
optimization can be achieved by choosing $z^*=z^*_{\rm opt}$ at which  
$\delta\sigma_{\alpha\beta}$ becomes minimum, so that the corresponding 
sensitivity has a maximum. The numerical results, reported in Tables~1 and 2, 
show that such optimization can allow a substantial increase of the lower 
bounds on $\Lambda_{RL}$ and $\Lambda_{LR}$, and a modest improvement for 
$\Lambda_{LL}$ and $\Lambda_{RR}$. 

In conclusion, the measurement of helicity amplitudes of process 
(\ref{eq:proc}) at the LC by means of suitable polarized integrated 
observables and 
optimal kinematical cuts to increase the sensitivity, 
would allow model-independent tests of four-fermion contact interactions, in 
particular as regards their chiral structure, up to mass scales 
$\Lambda_{\alpha\beta}$ of the order of 
40-100 times the C.M. energy, depending on the final fermion flavor and the 
degree of initial polarization. Work is in progress to assess the further 
increase in sensitivity to the new interactions which can be obtained if also 
a significant positron-beam polarization were available.

\end{document}